\documentclass[aps,superscriptaddress,prb,amsmath,amssymb,reprint,longbibliography]{revtex4-2}

\usepackage[USenglish]{babel}
\usepackage[T1]{fontenc}
\usepackage[utf8]{inputenc}
\usepackage{graphicx}
\usepackage{upgreek}
\usepackage[unicode]{hyperref}
\usepackage{multirow}
\usepackage{verbatim} 
\usepackage[usenames, dvipsnames]{color} 
\usepackage{verbatim} 
\usepackage[usenames, dvipsnames]{color} 
\usepackage{soul}
\usepackage{fixltx2e}
\usepackage{booktabs} 
\usepackage{siunitx}  
\usepackage{titlesec}
\usepackage{xurl}
\usepackage{subcaption}
\usepackage{tikz}
\usepackage{tikz-3dplot}
\usetikzlibrary{decorations.pathreplacing,decorations.markings}
\usetikzlibrary{arrows.meta,calc,decorations.markings}
\usepackage{standalone}
\usepackage{ragged2e}

\titlespacing*{\section}{0pt}{12pt}{6pt} 

\begin{document}

\raggedbottom

\setlength{\bibsep}{6pt} 
	
\title{Spin properties in droplet epitaxy-grown telecom quantum dots}

\date{\today}

\author{Marius Cizauskas}
\email{email: marius.cizauskas@tu-dortmund.de}
\affiliation{Experimentelle Physik 2, Technische Universit\"at Dortmund, 44221 Dortmund, Germany}
\affiliation{School of Mathematical and Physical Sciences, University of Sheffield, Sheffield, UK}

\author{Elisa M.~Sala}
\affiliation{EPSRC National Epitaxy Facility, Department of Electronic and Electrical Engineering, University of Sheffield, Sheffield, UK}

\author{Jon Heffernan}
\affiliation{EPSRC National Epitaxy Facility, Department of Electronic and Electrical Engineering, University of Sheffield, Sheffield, UK}

\author{A. Mark Fox}
\affiliation{School of Mathematical and Physical Sciences, University of Sheffield, Sheffield, UK}

\author{Manfred Bayer}
\affiliation{Experimentelle Physik 2, Technische Universit\"at Dortmund, 44221 Dortmund, Germany}

\author{Alex Greilich}
\email{email: alex.greilich@tu-dortmund.de}
\affiliation{Experimentelle Physik 2, Technische Universit\"at Dortmund, 44221 Dortmund, Germany}

\begin{abstract}
We investigate the spin properties of InAs/InGaAs/InP quantum dots grown by metalorganic vapor-phase epitaxy (MOVPE) deposition using droplet epitaxy, which emit in the telecom C-band. Using pump-probe Faraday ellipticity measurements, we determine electron and hole $g$-factors of $|g_e| = 0.934$ and $|g_h| = 0.471$, with the electron $g$-factor being nearly twice as low as typical molecular beam epitaxy Stranski-Krastanov (SK) grown samples. Most significantly, we measure a longitudinal spin relaxation time $T_1 = 2.95\si{\micro\second}$, representing an order of magnitude improvement over comparable MBE SK grown samples. Despite significant electron $g$-factor anisotropy, we observed that it is reduced relative to similar material composition samples grown with MBE or MOVPE SK methods. We attribute these g-factor anisotropy and spin lifetime improvements to the enhanced structural symmetry achieved via MOVPE droplet epitaxy, which mitigates the inherent structural asymmetry in strain-driven growth approaches for InAs/InP quantum dots. These results demonstrate that MOVPE droplet epitaxy-grown InAs/InGaAs/InP quantum dots exhibit favorable spin properties for potential implementation in quantum information applications.
\end{abstract}


\maketitle

\section{Introduction}

With the rapidly growing field of quantum computing, there is an increasing need of effective quantum network solutions. Currently, one of the significant problems in quantum networking is the decoherence of entangled states. When traveling a longer distance through a fiber, entangled states can get entangled with the environment, causing decoherence of the initial states~\cite{van_Enk_2001}. One solution is satellite-based communication, which would allow for transmission distances of over 800\,km~\cite{Rarity_2002, Bourgoin_2014} across a wide range of wavelengths, spanning from visible light to telecom C band~\cite{Giggenbach_2022}. 

However, the preferred option is to use the already established telecommunication fiber network. In this case, quantum repeaters are needed to mitigate against loss. Unlike classic repeaters, quantum repeaters do not amplify the signal; rather they perform entanglement swapping at dedicated nodes~\cite{howe_2024}. To implement quantum repeaters effectively, quantum storage devices that can maintain coherence for an extended period are required to allow the stored quantum state to become entangled with a photon~\cite{Bhaskar_2020}. Telecom quantum dots (QDs) are a good contender due to their efficient coupling to single-mode fibers (with additional nanostructures designed to couple to fibers)~\cite{Bauer_2021, Rahaman_2024}. It has been demonstrated that molecular beam epitaxy-grown InAs/InAlGaAs quantum dots can have hole carrier coherence times of up to 0.4\,\si{\micro\second}~\cite{Evers_2024}.

Better coherence times can also be achieved by exploring different methods besides strain-driven QD growth modes, most often Stranski-Krastanov (SK) growth in the context of QDs. Using SK poses a challenge growing InAs dots on InP, since anisotropic indium diffusion causes these quantum dots to develop highly elongated shapes~\cite{Gaweczyk_2022}. In contrast, quantum dots grown via metalorganic vapor-phase epitaxy deposition (MOVPE) droplet epitaxy (DE) exhibit enhanced structural symmetry~\cite{Huwer_2017}, resulting in longer optical coherence times~\cite{Sala_2023}. Consequently, MOVPE droplet epitaxy grown InP quantum dots present more favorable characteristics for implementation in quantum repeater applications.

Telecom quantum dots have been successfully grown via MOVPE using both self-assembly techniques~\cite{Miyazawa_2016, Pawel_2021} and droplet epitaxy (DE)~\cite{Laccotripes_2023, Anderson_2021, Sala_2020} methodologies. InAs/InP QDs demonstrate exceptional single-photon purity~\cite{Miyazawa_2016, Pawel_2021} at operational temperatures reaching 80\,K~\cite{Pawel_2021}, making them promising candidates for quantum optical applications. Comparative analyses indicate that MOVPE DE-fabricated InAs/InP quantum dots exhibit significantly enhanced coherence times relative to strain-based growth methods~\cite{Anderson_2021}. Additionally, experiments have shown the feasibility of quantum entanglement generation using MOVPE DE-grown structures~\cite{Laccotripes_2023}, thereby expanding their potential use in quantum communication protocols. This paper aims to characterize the spin properties of InAs/InGaAs/InP quantum dots produced through MOVPE droplet epitaxy using a pump-probe measurement scheme to evaluate their potential applications, and conduct comparative analyses with samples grown by other techniques.

\section{Experimental details}

\begin{figure*}[]
\centering
\includegraphics[width=12cm]{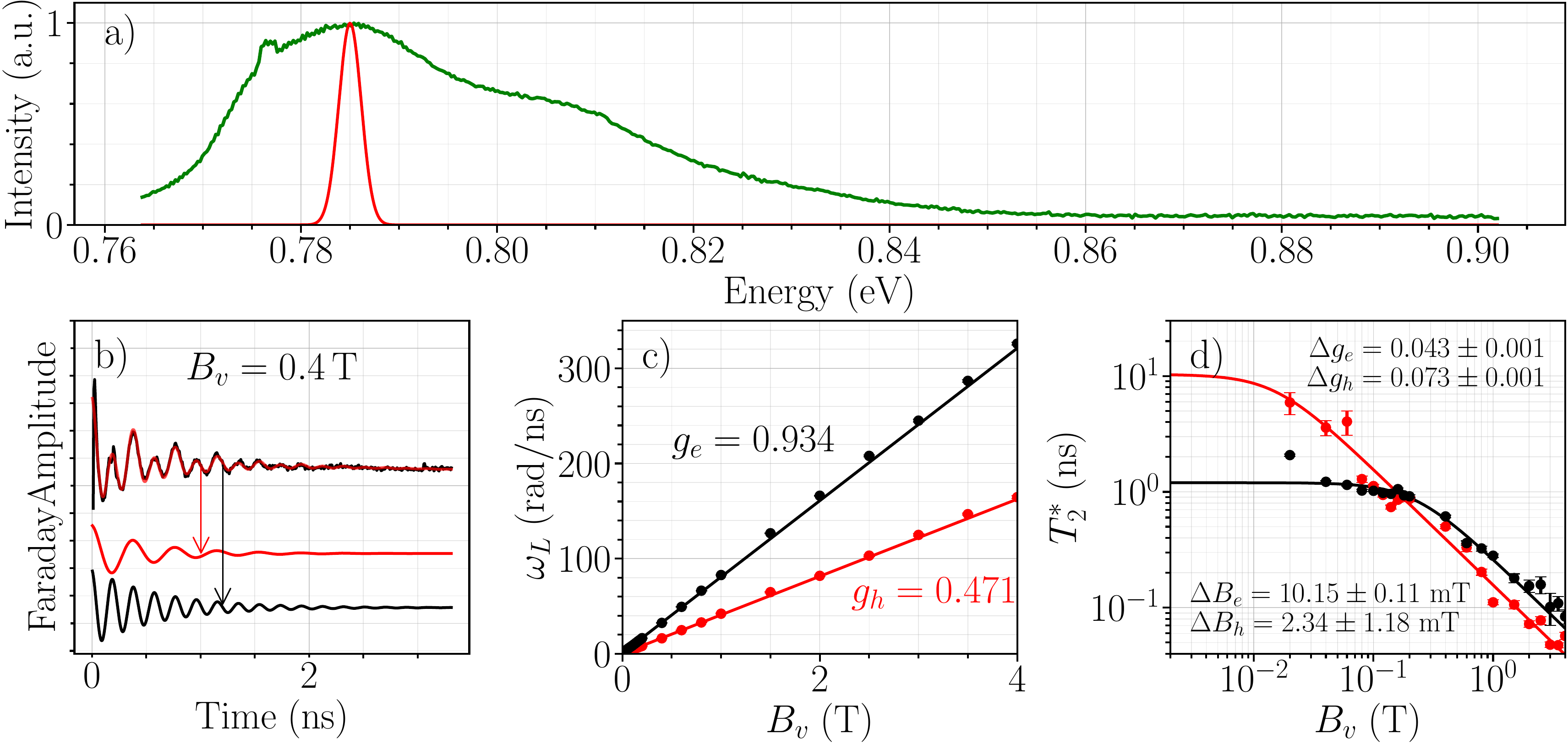}
\caption{\justifying MOVPE droplet epitaxy grown InAs/InGaAs QD spin properties of carriers in transverse magnetic field. Excitation energy for all measurements is 0.785\,eV. Measurements are done at 6K temperature, 20 mW pump power and 2 mW probe power. (a) Photoluminescence of the sample (green) and the excitation laser spectrum (red). (b) QD oscillatory behavior is displayed when a transverse magnetic $B_V = 0.4$\,T is applied and the Faraday ellipticity signal is measured. The black curve is the raw data, and the red curve is the fit. The red and black curves correspond to hole and electron contributions, respectively. (c) Larmor precession frequencies depending on the transverse magnetic field for electron (black) and hole (red). The $g$ factor calculated for both carriers is displayed in the plot. (d) Spin dephasing times as a function of transverse magnetic field. The calculated $g$-factor and $B$ field spreads for each component are displayed in the top right.}
\label{fig:one}
\end{figure*}

\begin{figure}[h]
    \centering
    \begin{tikzpicture}[scale=1]
    \definecolor{inp}{RGB}{70,130,180}
    \definecolor{ingaas}{RGB}{255,165,0}
    \definecolor{inas}{RGB}{220,20,60}
    
    \fill[inp] (0,0) rectangle (8,1.5);
    \node[right, font=\Large] at (0.2,0.75) {InP Substrate};
    
    \draw[thick] (0,1.5) -- (8,1.5);
    
    \fill[inp] (0,1.5) rectangle (8,3.5);
    \node[right, font=\Large] at (0.2,2.5) {Layer 1: InP Buffer (300 nm)};
    
    \fill[ingaas] (0,3.5) rectangle (8,4.2);      
    \fill[inas] (0,4.2) rectangle (8,4.9);        
    \fill[inp!80] (0,4.9) rectangle (8,6.1);      
    \fill[inp] (0,6.1) rectangle (8,8.5);       
    
    \node[right, font=\Large] at (0.2,3.85) {Layer 2: Ga\textsubscript{0.47}In\textsubscript{0.53}As (5.0 nm)};
    \node[right, font=\Large] at (0.2,4.55) {Layer 3: InAs QDs (2.0 nm)};
    \node[right, font=\Large] at (0.2,5.5) {Layer 4: InP Capping (20.0 nm)};
    \node[right, font=\Large] at (0.2,7.3) {Layer 5: InP Burial (80.0 nm)};
    
    \draw[decoration={brace,amplitude=5pt},decorate] (-0.4,3.5) -- (-0.4,8.5);
    \node[left, font=\Large, rotate=90] at (-0.8,7.3) {Repeat 3x};
    
    
        
        
        
        
    
\end{tikzpicture}
    \caption{\justifying{Layer structure of MOVPE droplet epitaxy grown InAs/InGaAs/InP quantum dots showing the substrate, buffer layer, and three repetitions of the quantum dot active region consisting of InGaAs interlayer, InAs QDs, InP capping, and InP burial layers.}}
    \label{fig:sample}
\end{figure}
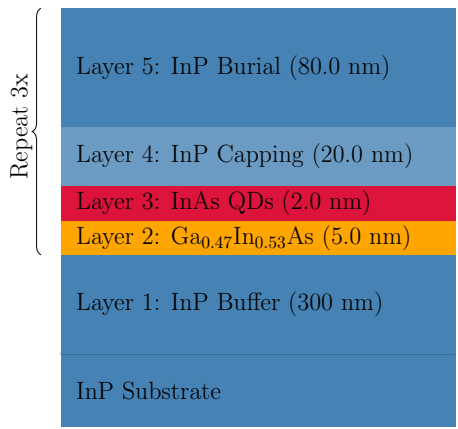

The QD sample under study is grown by MOVPE on a (100)-oriented InP substrate in an Aixtron CCS 3x2" reactor. The sample's layer structure is presented in Figure~\ref{fig:sample}. None of the layers are intentionally doped. The QD density is approximately $\mathrm{10^{10}\ cm^{-2}}$. Measurements are performed on the central part of the wafer. The signal quality exhibits strong spatial dependence within the sample, possibly due to inhomogeneous QD distribution. For more details on the growth process, see Ref.~\cite{Sala_2022}.

The sample is placed in a helium bath cryostat with an electronic heater to control the temperature. A superconducting magnet inside the cryostat is used to apply magnetic fields. The magnetic field can be applied either longitudinally to the $\textbf{k}$-wavevector of light (Faraday geometry) or transverse (Voigt geometry) by rotating the sample relative to the magnet. 


The sample spin dynamics are measured using a pump-probe method. The laser source is a Ti:Sapphire pulsed laser at 76.7\,MHz repetition rate, pumping an optical parametric oscillator, which produces 2\,\si{\nano\meter} wide 1.8 ps duration pulses with variable wavelength in the telecom C or L-band. The pulses are split into two paths, pump and probe, where the probe path goes through a mechanized delay line for scanning delay. The beams are then focused to a diameter of 100\,\si{\micro\meter} for the probe and 150\,\si{\micro\meter} for the pump and overlapped at the sample.

The pump pulses are modulated between $\sigma^+$ and $\sigma^-$ using an electro-optic modulator with a frequency range of 1\,kHz to 2\,MHz (10\,kHz for all Voigt geometry measurements), while the probe pulses are modulated in intensity using a photoelastic modulator at 86\,kHz followed by a Glan-Thompson polarizer. The circularly polarized pump pulses create the spin-polarized charge carriers in the sample, while the linearly polarized probe pulses are used to measure the spin polarization by the Faraday effect~\cite{Yugova_2009}. 
To reduce the possible amount of scattered pump light in the probe detection channel, a 10\,\si{\micro\meter} pinhole was used to let through only the probe pulses. It is followed by a quarter-wave plate converting the circular components of the induced ellipticity into linear ones, a Wollaston polarizer separating the linearly polarized probe into two orthogonal linearly polarized components, which are then directed to the balanced photodiodes. The probe signal is demodulated at the difference modulation frequency and amplified by a lock-in amplifier.

\section{Experimental results}

\begin{figure}[]
\includegraphics[width=\columnwidth]{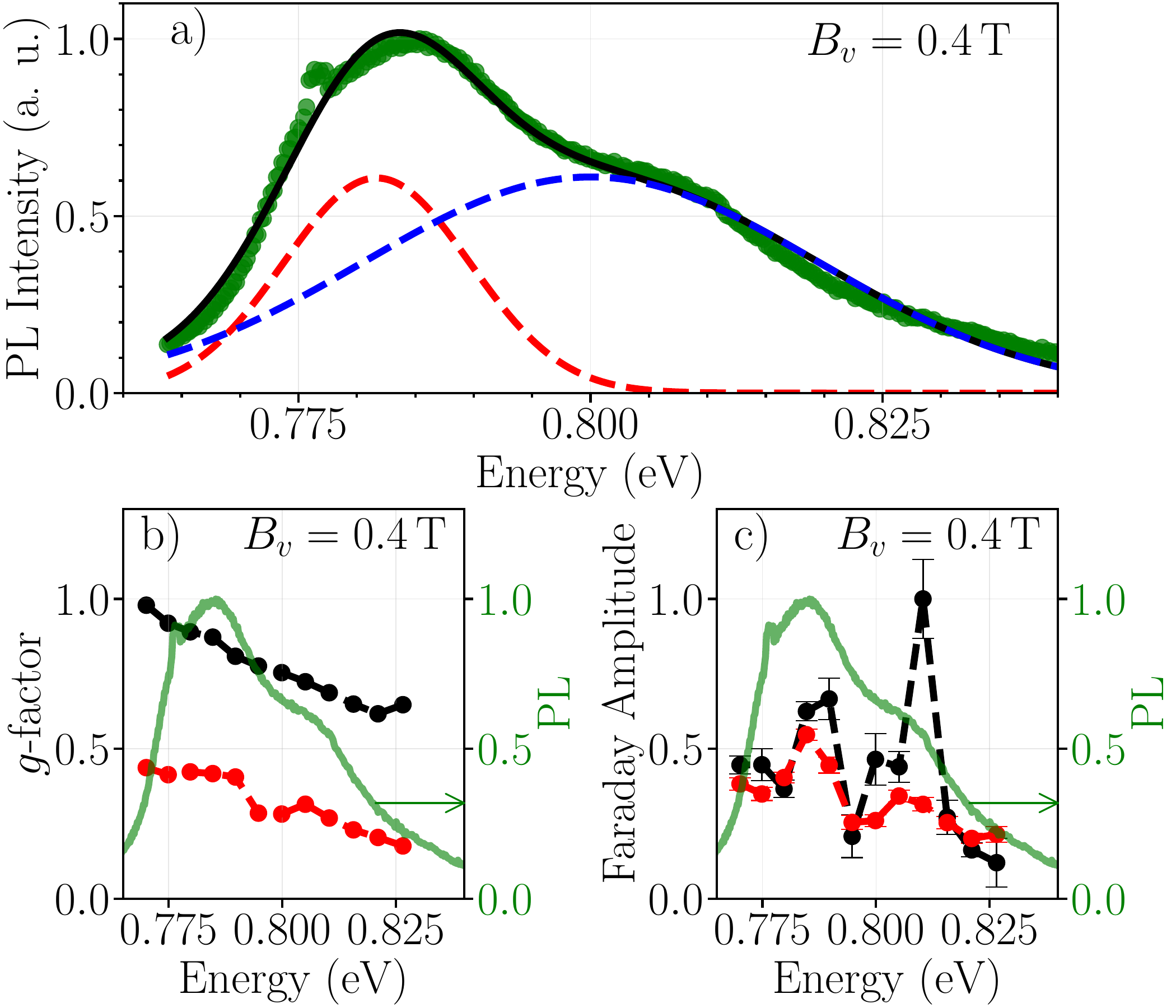}
\caption{\justifying PL spectra, spectral dependence of $g$-factors and Faraday ellipticity amplitudes at $B_V = 0.4$\,T of electron and hole components in MOVPE droplet epitaxy grown InAs/InGaAs QDs. Measurements are done at 6 K temperature, 20 mW pump power and 2 mW probe power. (a) PL spectrum of the sample (green markers) fitted with a bimodal distribution (black curve, with red and blue dashed curves showing components of the bimodal). (b) Absolute values of $g$-factor as a function of excitation energy for electron (black) and hole (red) components. The normalized PL spectrum is shown as a green line (plotted on the right axis). (c) Normalized Faraday ellipticity amplitudes for electrons (black) and holes (red) as a function of excitation energy.}
\label{fig:two}
\end{figure}

Figure~\ref{fig:one}a shows photoluminescence (PL) of the sample at 6K, which was obtained by exciting the sample with a laser diode at 2.33 eV photon energy. The pump excitation is done at the peak of the PL 0.785\,eV (shown in Fig.~\ref{fig:one}a by a red Lorentzian) for two reason: to ensure that the smallest possible distribution of QD sizes is excited and because signal amplitude was highest, shown in Fig.~\ref{fig:two}c. There is a larger amplitude peak at higher energies, however, at that point a larger distribution of dots would be excited, which would degrade the measured optical properties of the sample.

Figure~\ref{fig:one}b displays QD spin polarization signal, measured by Faraday ellipticity, with an applied transverse magnetic field of 0.4\,T. Measurements are done at 0.4\,T since at this magnetic field strength the signal provides optimal signal clarity. Upon optical excitation, an electron-hole pair is formed, which results in two oscillatory components, one for the hole and one for the electron. The exponential decay is spin dephasing, which is influenced by the inhomogeneity of the excited QD ensemble. The Faraday ellipticity signal can therefore be fitted with the following equation:

\begin{equation}
S = \sum_i A_i \cos(\omega_i t)\exp\left(-\frac{t^2}{T_{2,i}^{*2}}\right),
\label{eq:1}
\end{equation}

\noindent where $i$ -- index for component, $A_i$ -- amplitude of component, $\omega_i$ -- component Larmor precession frequency, and $T^{*2}_{2,i}$ -- component spin dephasing time. The $g$-factor of both can be calculated by following a linear relationship of Larmor precession frequency on the strength of the applied transverse magnetic field:

\begin{equation}
\omega_L = \frac{g\mu_B B_V}{\hbar},
\label{eq:2}
\end{equation}

\noindent where $\omega_L$ -- Larmor precession frequency, $\mu_B$ -- Bohr magneton constant, $B_V$ -- transverse magnetic field strength, and $\hbar$ -- reduced Planck constant. Figure~\ref{fig:one}c shows the frequency dependence on the transverse magnetic field strength. A linear equation is observed for both hole and electron, from which the $g$-factors can be extracted using Eq.~\ref{eq:2}. We obtain $|g_e| = 0.934$ and $|g_h| = 0.471$. It should be noted that the pump-probe measurement method that we are using can only be used to obtain absolute g-factor values. Therefore, the g-factor values obtained here and all subsequent ones are absolute. 

Figure~\ref{fig:one}d displays the extracted spin dephasing times depending on the transverse magnetic field. Spin dephasing is affected mostly by $g$-factor spread due to QD size and content inhomogeneity and by hyperfine coupling to the nuclear spins. The relationship between $T_2^*$ and $B_V$ is described by~\cite{Evers_2024}:

\begin{equation}
T^*_{2,i} = \hbar/\sqrt{(\Delta g_i \mu_B B_V)^2 + (g_i \mu_B \Delta B_i)^2},
\label{eq:3}
\end{equation}

\noindent where $\Delta g_i$ -- $g$-factor spread and $\Delta B_i$ -- magnetic field fluctuation spread, which quantifies the fluctuation of the nuclear spins. Figure~\ref{fig:one}d shows data which is fitted by Eq.~\ref{eq:3}. At magnetic fields below 0.1\,T, the dephasing times are limited by the hyperfine interaction to nuclear spins. The data show that spin dephasing at low magnetic field is saturated at $T^{*}_{2,e} = 1.2$\,ns for electrons and at $T^{*}_{2,h} = 10.0$\,ns for holes. This is expected due to the nuclear interaction hyperfine constant of a hole being a magnitude smaller than for the electron~\cite{Testelin_2009}. This is further supported by the magnetic field spread values seen from the fit, which are $\Delta B_e = 10.15 \pm 0.11\,$mT and $\Delta B_h = 2.34 \pm 1.18\,$mT. At higher magnetic fields, contributions from $g$-factor spread become significant, causing the spin dephasing time to reduce. The $g$-factor spread values are $\Delta g_e = 0.043 \pm 0.001$ and $\Delta g_h = 0.073 \pm 0.001$. It is seen that the electron $g$-factor spread is lower than that of the hole, which is expected, due to the smaller effective mass for the electron and the correspondingly broader wavefunction, leading to the spatial averaging of the inhomogeneities. For comparison, in the case of a coupled multi-layer MBE SK grown InAs QDs, despite the electron having a broader wavefunction, the $g$-factor spread of electrons was higher than that of holes. This is attributed to a higher probability of electron tunneling between the layers in the structure, due to its lower effective mass, providing a bigger variation of environment, causing increased g-factor spread~\cite{Evers_2024}. In our sample, tunneling probability is greatly reduced due to barrier thickness of 100 nm, as opposed to 15 nm in Ref.~\cite{Evers_2024}.

The $g$-factor spread values are also lower for both components compared to the MBE SK sample, which is possibly caused by MOVPE DE grown samples having reduced QD size inhomogeneity. Figure~\ref{fig:two} shows the PL of the measured sample with a bimodal distribution fit. Compared to the previously mentioned MBE SK samples, the spectrum appears to be similar ~\cite{Evers_2024, Belykh_2015, Carmesin_2017}. It has been previously demonstrated that MOVPE for deposition results in a bimodal distribution of QD sizes~\cite{Sala_2022, Banfi_2025}. This is due to the fact that the smaller QDs grow first and then merge together to form larger ones~\cite{Sala_2022, Banfi_2025}. It may also be a consequence of stacking multiple layers of QDs, where a different layer may have a different distribution of QD sizes. Two different QD distributions can be further confirmed by looking at Fig.~\ref{fig:two}c, which shows Faraday amplitude of carrier components measured at different excitation energies. It can be seen that there are two peak values for both electron and hole components.

Referring back to results in Fig.~\ref{fig:one}c, if one compares our results to the grown InAs/InAlGaAs monolayer QD sample in Ref.~\cite{Belykh_2015}, it shows more similar $g$-factor spread values of $\Delta g_e = 0.02$ and $\Delta g_h = 0.05$. Our sample exhibits a PL spectral width of approximately 19 meV for the narrower size distribution seen in Fig.~\ref{fig:two}a red curve and approximately 46 meV for the wider size distribution shown by the blue curve. The sample in Ref.~\cite{Belykh_2015} shows PL width of 60 meV, which is similar to our case if we combine the widths of both components, resulting in a width of 65 meV. Since the width of our PL is larger, it is expected that our g-factor spread values are also marginally larger.

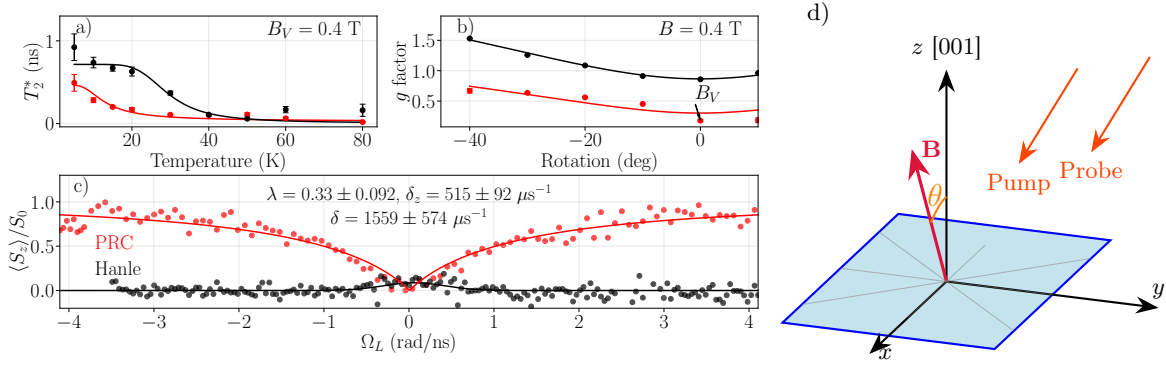
\begin{figure*}[]
\centering
\begin{minipage}{0.55\textwidth}
    \centering
    \includegraphics[width=10cm]{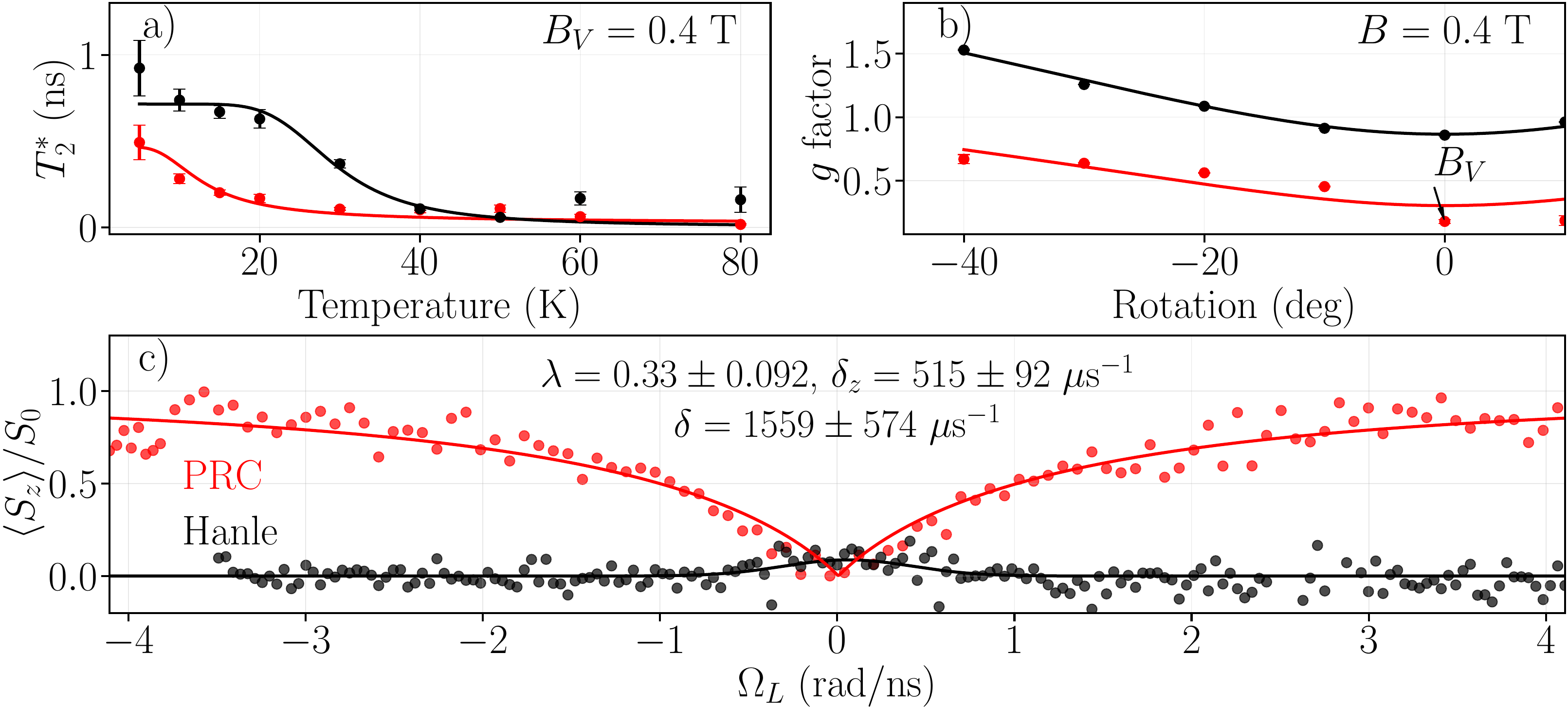}
\end{minipage}
\begin{minipage}{0.3\textwidth}
    \llap{\text{d)}\hspace{-3em}}
\tdplotsetmaincoords{70}{110}
\begin{tikzpicture}[tdplot_main_coords, scale=1.5]

\definecolor{samplecolor}{RGB}{173,216,230}
\definecolor{bfieldcolor}{RGB}{220,20,60}
\definecolor{beamcolor}{RGB}{255,69,0}

\filldraw[samplecolor, opacity=0.7, thick] 
    (-1.5,-1,0) -- (1.5,-1,0) -- (1.5,1,0) -- (-1.5,1,0) -- cycle;
\draw[blue, thick] (-1.5,-1,0) -- (1.5,-1,0) -- (1.5,1,0) -- (-1.5,1,0) -- cycle;

\draw[-{Stealth[length=3mm]}, thick] (0,0,0) -- (2,0,0) node[right] {$x$};
\draw[-{Stealth[length=3mm]}, thick] (0,0,0) -- (0,2,0) node[above] {$y$};
\draw[-{Stealth[length=3mm]}, thick] (0,0,0) -- (0,0,2) node[above] {$z$ [001]};

\def\btheta{30} 
\def\bmag{1.8}  
\pgfmathsetmacro{\bx}{\bmag*sin(\btheta)}
\pgfmathsetmacro{\bz}{\bmag*cos(\btheta)}

\draw[-{Stealth[length=4mm]}, bfieldcolor, very thick] 
    (0,0,0) -- (\bx,0,\bz) node[right] {\textbf{B}};

\tdplotsetrotatedcoords{0}{0}{0}
\draw[orange, thick] 
    plot[domain=90:60, smooth] ({0.8*cos(\x)}, {0}, {0.8*sin(\x)});
\node[orange, font=\large] at (1.8,0.55,1.5) {$\theta$};

\draw[-{Stealth[length=3mm]}, beamcolor, thick] 
    (-1.7,1.3,1.7) -- (-1,1,1.0) node[below] {\small Probe};
\draw[-{Stealth[length=3mm]}, beamcolor, thick] 
    (-1.2,0.8,1.7) -- (-0.5,0.5,1.0) node[below] {\small Pump};

\draw[gray, opacity=0.5] (-1,-1,0) -- (1,1,0);
\draw[gray, opacity=0.5] (-1,1,0) -- (1,-1,0);
\draw[gray, opacity=0.5] (0,-1,0) -- (0,1,0);
\draw[gray, opacity=0.5] (-1,0,0) -- (1,0,0);

\end{tikzpicture}
\end{minipage}
\caption{\justifying a) Temperature dependence of spin dephasing times at a transverse magnetic field of 0.4\,T for electrons (black) and holes (red), fitted with Arrhenius equation. b) Anisotropy of $g$-factors measured at $B = 0.4$\,T as a function of magnetic field rotation angle. c) Magnetic field dependence of electron spin polarization showing the polarization recovery curve (red) and Hanle effect (black). Measurements are done at 6K temperature, 2 mW probe power, 17 mW pump power for PRC and 15 mW for Hanle. d) Sample rotation geometry showing magnetic field B at angle $\theta$ relative to the [001] growth axis, enabling measurement of g-factor anisotropy between transverse and parallel orientations.}
\label{fig:three}
\end{figure*}

The emission energy of a QD is dependent on its size, meaning that the $g$-factors of components change due to different size QDs being excited at different energies~\cite{Kleemans_2009, Sheng_2007}. Generally, electron $g$-factor decreases and hole $g$-factor increases as the excitation energy increases~\cite{Evers_2024, Belykh_2015}. In Fig.~\ref{fig:two}b it can be seen, that the electron $g$-factor values follow what is seen in other QD samples, but for the hole it differs. The explanation for that is that the magnetic field was not exactly along [001] direction ($\theta = 0$ in Fig.~\ref{fig:three}d). It has been theoretically shown that hole $g$-factor demonstrates the same behavior as in our case when the magnetic field is applied along [011] direction~\cite{Pryor_2006}. S.~A.~Crooker demonstrates this experimentally with InGaAs/GaAs QDs~\cite{Crooker_2010}, where by applying a magnetic field along the [0\textrm{\=1}1] instead of [011], the hole $g$-factor gradient inverts, which is in good agreement to our observed hole behavior.

Figure~\ref{fig:three}a shows how the spin dephasing of electron and hole carriers changes as temperature changes. These results further confirm that we observe both an electron and a hole component in the previous measurements. It can be seen from the data that the dephasing time starts decaying for one of the components at 10\,K (holes) and for the other at 20\,K (electrons). The data was fitted with Arrhenius equation~\cite{Greilich_2012}:

\begin{equation}
    \frac{1}{T_2^*} = \frac{1}{T_g} + \frac{1}{T_e}\exp\left[-\frac{E_a}{k_BT}\right],
    \label{eq:four}
\end{equation}

\noindent where $T_g$ -- ground state relaxation time, $T_e$ -- excited state relaxation time and $E_a$ -- activation energy. The fit gives activation energies of 15\,meV for electrons and 3.6\,meV for holes.

It has been demonstrated in 900 nm (In,Ga)As/GaAs QDs that spin coherence for holes starts decaying at 10\,K and 20\,K for electrons, since as temperature rises, elastic scattering processes that leverage the broad phonon sidebands are expected to significantly accelerate the deterioration of hole spin coherence~\cite{Varwig_2013}. However, while there is good agreement between our sample and the example in Ref.~\cite{Varwig_2013} in terms of when dephasing starts decaying, spin dephasing is only bounded by coherence times. Coherence does not have a direct influence on dephasing, and in the previously shown example, the coherence times are significantly larger than the dephasing times displayed in Fig.~\ref{fig:three}. A more relevant reference~\cite{Mikhailov_2018} shows spin dephasing times for both electrons and holes in InAs/InAlGaAs/InP QDs. It is demonstrated that electron dephasing starts decaying at higher temperatures compared to hole, which is the same behavior we see in our sample. This behavior is due to the inherently stronger spin-orbit interactions in the valence band, which enhance hole coupling to phonons~\cite{Li_2020}. The biggest difference is the activation energy, which is significantly larger in Ref.~\cite{Mikhailov_2018}. This is likely due to the Ref.~\cite{Mikhailov_2018} QDs having greater confinement, where it has been shown that the presence an AlGaAs barrier increases activation energy~\cite{Das_2017}.

Figure~\ref{fig:three}b shows how the $g$-factor of electron and hole change as the magnetic field is rotated; in this case, the magnetic field direction was kept constant, and the sample was rotated. The data is fitted with the following equation \cite{Belykh_2016_anisotropy}:

\begin{equation}
    \omega_{\mathrm{L}}\left(\theta\right) = \sqrt{(g_{\perp}\cos\theta)^{2} + (g_{\parallel}\sin\theta)^{2}},
    \label{eq:five}
\end{equation}

\noindent where $\theta$ -- magnetic field at angle relative to the [001] growth axis, $g_{\perp}$ -- transverse magnetic field $g$-factor and $g_{\parallel}$ -- parallel magnetic field $g$-factor. The transverse $g$-factors were $g_{\perp,e} = 0.87 \pm 0.017, \ g_{\perp,h} = 0.30 \pm 0.087$ and parallel $g_{\parallel,e} = 2.1 \pm 0.043, \ g_{\parallel,h} = 1.1 \pm 0.16$. Compared to another InAs/InAlGaAs sample~\cite{Belykh_2016_anisotropy}, the results are not as expected, where in our case the electron anisotropy is larger than the hole anisotropy. However, in another InAs/InP sample, it is shown that a decrease in excitation energy (from 0.91\,eV to 0.86\,eV) causes a significant increase in electron $g$-factor anisotropy~\cite{Belykh_2016}. Measurements for the electron $g$-factor anisotropy are also done in Ref.~\cite{Belykh_2016_anisotropy}. However, they show barely any change. This indicates that the presence of InP impacts the electron $g$-factor anisotropy. The possible cause of this is the increasing asymmetry of the QDs as the size increases~\cite{Belykh_2016}. It is also known that due to the low lattice mismatch between InAs and InP (3\%), both MBE and MOVPE strain-based growth methods for such QDs have an asymmetric growth profile~\cite{Yan_2024}, which could be further exacerbated by larger size of the QDs. This would explain the increasing anisotropy for the mentioned InAs/InP sample, which was grown by MOVPE SK method. Based on this trend, our sample should exhibit even greater anisotropy since our excitation energy is 0.075 eV lower than the highest value in Ref.~\cite{Belykh_2016}, where they already demonstrated significant anisotropy increases with decrease of excitation energy. However, our measured electron g-factor anisotropy is comparable, rather than larger for both MOVPE and MBE SK grown InAs/InP QDs. It has been shown that the same QDs can be produced with better symmetry properties by using MOVPE droplet epitaxy~\cite{Holewa_2021}, which mitigates the expected anisotropy increase in our sample despite the lower excitation energy in comparison to Ref.~\cite{Belykh_2016}.

\begin{figure*}[]
\centering
\includegraphics[width=12cm]{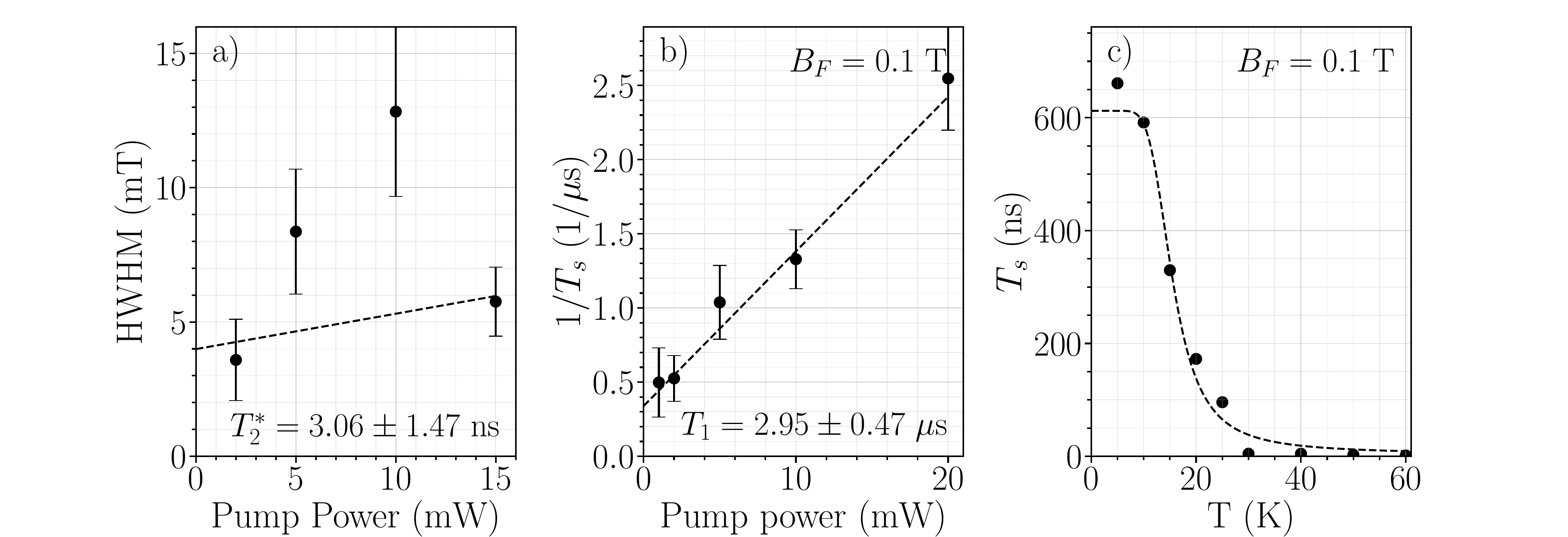}
\caption{\justifying Spin dynamics measurements of quantum dots showing: (a) Half-width at half maximum (HWHM) of the Hanle curve as a function of pump power, showing electron spin lifetime at zero power extrapolation. (b) Inverse electron spin lifetime versus pump power measured with a $B_F=0.1$\,T parallel magnetic field applied, with linear extrapolation to zero to obtain electron longitudinal spin relaxation time $T_1$. (c) Temperature dependence of electron spin lifetime with a 0.1 T parallel magnetic field applied, fitted with an Arrhenius equation (dashed line).}
\label{fig:four}
\end{figure*}

Figure~\ref{fig:three}c shows the relationship between the polarization recovery curve (PRC) and Hanle curve. Both PRC and Hanle curve were obtained by sweeping magnetic field strength and measuring Faraday ellipticity. For PRC, the magnetic field was parallel, and for Hanle, it was transverse. The PRC and Hanle curve can be fitted with the following anisotropic hyperfine interaction model \cite{Smirnov_2020}:

\begin{equation}
H(\Omega_L) = \lambda^2 [\ln(2/\lambda) - 1] \exp\left(-\frac{\Omega_L^2}{\delta_z^2}\right) \ll 1
\label{eq:H_function_exact}
\end{equation}

\begin{align}
P(\Omega_L) &= \frac{|\Omega_L|}{\delta}\left\{\sin\left(\frac{|\Omega_L|}{\delta}\right)\text{Ci}\left(\frac{|\Omega_L|}{\delta}\right) \right. \nonumber \\
&\quad + \left. \cos\left(\frac{|\Omega_L|}{\delta}\right)\left[\frac{\pi}{2} - \text{Si}\left(\frac{|\Omega_L|}{\delta}\right)\right]\right\},
\label{eq:P_function_exact}
\end{align}


\noindent where $P$ -- PRC curve, $H$ -- Hanle curve, $\delta_z$ -- dispersion of the nuclear fluctuations in z direction, $\delta$ -- dispersion of nuclear fluctuations in (xy) directions, $\lambda$ -- anisotropy parameter, which relates dispersions $\delta_z = \lambda \delta$, $\Omega_L = g_e \mu_B B/\hbar$, Ci -- cosine integral function and Si -- sine integral function. This results in anisotropy parameter $\lambda = 0.33 \pm 0.092$, dispersion $\delta_z = 515 \pm 92\,\mu s^{-1}$ for the z direction and $\delta = 1559 \pm 574\,\mu s^{-1}$ for the (xy) directions. The PRC being relatively wide and V-shaped indicates that it is the spin dynamics of the electron that are being measured~\cite{Zhukov_2018}. The implications of these parameters will be discussed in the discussion part.

Lastly, Fig.~\ref{fig:four} shows the different spin times of our sample. Figure~\ref{fig:four}a is measured by performing Hanle measurements at different pump powers. The HWHM from the Hanle curves can then be extracted, and the intrinsic spin lifetime can be calculated by extrapolating to zero pump power and applying the following equation~\cite{Rittman_2022}:

\begin{equation}
T_s = \frac{\hbar}{g_e\mu_B B_{\text{HWHM}}}.
\label{eq:T_s}
\end{equation}

\noindent This results in a spin dephasing time of $T_2^* = 3.06 \pm 1.47 \ ns$. In Fig.~\ref{fig:four}b $T_1$ is measured by using the spin inertia method~\cite{Heisterkamp_2015}. A longitudinal magnetic field is applied to decouple the carrier spins from the fluctuations of the nuclear spins. Fixing the delay at a negative time delay, the modulation frequency is swept, and the Faraday ellipticity signal is measured. When the modulation period falls below the spin lifetime $T_s$, measured spin polarization decreases, causing the average signal amplitude to diminish. This can be described by the following equation \cite{Evers_2024}:

\begin{equation}
    S(f_m) = \frac{S_0}{\sqrt{1 + (2\pi f_m T_s)^2}},
    \label{eq:S}
\end{equation}

\noindent where $S_0$ -- amplitude of the Faraday ellipticity signal and $f_m$ -- EOM modulation frequency of the pump beam polarization. This allows the extraction of the spin lifetimes at different pump powers and obtaining the longitudinal spin relaxation by extrapolating to zero pump power, which is $T_1 = 2.95 \pm 0.47\,\mu s$. Finally, Fig.~\ref{fig:four}c shows the spin lifetime dependence on varying temperature. It is fitted with Arrhenius equation, which shows similar activation energy to the electron activation energy seen in Fig.~\ref{fig:three}a. This indicates that we are most likely measuring the spin lifetimes for electrons. In comparison, an MBE-grown InAs/InAlGaAs C-band QD sample shows $T_1 = 0.5\,\mu s$~\cite{Evers_2024} and another MBE-grown InAs/InAlGaAs/InP C-band QD sample shows $T_1 = 0.2\,\mu s$~\cite{Mikhailov_2018}. Our sample shows an order-of-magnitude improvement, indicating that it is advantageous to use MOVPE DE for QD growth.

\section{Discussion}

Improvement in $T_1$ time is seen, an order of magnitude larger compared to other MBE SK grown telecom QD samples~\cite{Evers_2024, Mikhailov_2018}. However, there are 900 nm wavelength range QD samples that show $T_1$ times in the millisecond range~\cite{Kroutvar_2004}. One of the factors that influence $T_1$ times is the degree of confinement in the QD~\cite{Lawrie_2020}. In the case of Ref.~\cite{Kroutvar_2004} there are two factors that increase the confinement of their QDs: an asymmetric AlGaAs barrier, where it has been shown that barriers with AlGaAs tend to increase confinement~\cite{Das_2017}, and small QD size. In our sample, the barrier only consists of InP, and the size is larger, causing reduced confinement energy. Another contributing factor is the hyperfine interaction with nuclear spins. It has been demonstrated in a p-doped InAs sample that the hole $T_1$ time is limited to 1\,\si{\micro\second} due to coupling to nuclear spins~\cite{Fras_2012}. In our case, we are measuring the electron $T_1$, however, as we have shown with Fig.~\ref{fig:three}c results, in our sample, there is significant hyperfine interaction anisotropy, which reduces $T_1$~\cite{Smirnov_2020, Mikhailov_2018}.

To quantitatively analyze this hyperfine interaction anisotropy observed in Fig.\ref{fig:three}c, we first consider the theoretical framework. In an ideal case, the data would be fit using a basic model \cite{Smirnov_2020}: 

\begin{equation}
H(\Omega_L) \approx \frac{2}{3} \frac{\delta^2}{2\delta^2 + \Omega_L^2}
\label{eq:H_function}
\end{equation}

\begin{equation}
P(\Omega_L) \approx \frac{1}{3} \frac{2\delta^2 + 3\Omega_L^2}{2\delta^2 + \Omega_L^2},
\label{eq:P_function}
\end{equation}

\noindent but this model provides an incomplete description of the system, as the observed ratio between PRC and Hanle curves deviates from the expected 2:1 relationship. This basic model assumes that the distribution function of the nuclear spin precession frequency is isotropic. It also assumes that the nuclear spins on the timescale of electron precession are effectively static. These assumptions would result in the 2:1 distribution of PRC and Hanle as seen in Eqs.~\ref{eq:H_function} and \ref{eq:P_function}. However, in our case a 2:1 distribution is not seen. Section 3 of Ref.~\cite{Smirnov_2020} provides a different model which still assumes that the nuclear spins are static, but that their field fluctuations are anisotropic. The model with hyperfine interaction contributions is seen in Eqs. \ref{eq:H_function_exact} and \ref{eq:P_function_exact}.

Fitting these functions for our data shows hyperfine interaction anisotropy of $\lambda = 0.33$, which indicates that for our sample, the nuclear field is distributed mostly in the (xy) plane, which results in an average spin polarization of zero, seen in our data in Fig.~\ref{fig:three}c. Our data is also similar to the case shown in Section 4 of Ref.~\cite{Smirnov_2020}, where the nuclear spins are no longer considered static and have a finite spin correlation time of $\tau_c$, and unrelated effects to hyperfine interaction (electron-phonon or spin-orbit interactions) decay the average spin during spin relaxation time, denoted as $\tau_s$. The similarity is seen in the case where $\tau_c \gg \tau_s$. However, the PRC Lorentzian is wider than in our case, and the value at $|\Omega_L| \gg 0$ is not 1. Similar experimental results for p-doped (In,Ga)As/GaAs QDs are presented in Section VII. D. of Ref.~\cite{Smirnov_2020}. The main difference is in the shape of the PRC curve, which in their case is composed of two Lorentzians, since both electrons and holes are present. Otherwise, it can be seen that their ratio is also not 2:1, which is attributed to the $g$-factor anisotropy. Our sample electron $g$-factor anisotropy (Fig.~\ref{fig:three}b) is larger than the anisotropy seen for the QD sample in Ref.~\cite{Smirnov_2020}, explaining why the hyperfine anisotropy is more pronounced for our sample. It is also seen when comparing the dispersion values, where in Ref. \cite{Smirnov_2020} they see $\delta = 440 \ \mu s^{-1}$ and for our sample $\delta = 1559 \ \mu s^{-1}$, which is more than twice as large. This is expected, since our electron g-factor anisotropy is also more than twice as large. This further confirms that both electron and hole charge carriers are present in the previous measurements, and demonstrates that the electron $g$-factor exhibits pronounced anisotropy.

Beyond the anisotropic characteristics discussed above, another important feature of our sample is the reduced electron $g$-factor magnitude. It should be said that the measurement method presented in Fig.~\ref{fig:one} does not allow to determine the signs of the $g$-factor, however, the absolute value is still significantly different compared to other telecom QD samples, which display absolute electron $g$-factors closer to 2~\cite{Belykh_2015, Evers_2024}. This is related to the increased electron $g$-factor due to the QD shape asymmetry. Ref.~\cite{Belykh_2016} demonstrates lower $g$-factors at higher excitation energies than in our case, which is attributed to the InAs/InP QDs having a structure similar to a flat disc~\cite{Banfi_2025}. Therefore, our MOVPE DE grown sample shows improvement in structural symmetry by having relatively better electron anisotropy properties and lower electron $g$-factor value, compared to MBE or MOVPE grown with strain methods InAs/InP quantum dots.

Additionally, the temperature dependence of $T_1$ in Fig.~\ref{fig:four}c is as expected. Assuming that at low temperatures the primary mechanism for spin relaxation is due to hyperfine interaction, we expect to see similar behavior for electron spin dephasing as in Fig.~\ref{fig:three}a, where the relaxation would remain constant and limited by hyperfine interaction and at higher temperatures it would decay due to phonon interactions. However, the relaxation time starts decaying earlier compared to spin dephasing temperature dependence. This may be caused by hyperfine coupling variations mediated by lattice phonons~\cite{Hernandez_2008}. It is mentioned in Ref.~\cite{Hernandez_2008}, that $T_1$ time is protected from the main inelastic-scattering mechanism due to confinement, however, as we have determined before, our sample's confinement is reduced, which is why it is possible that we are observing such temperature dependence.

The electron spin dephasing times at zero pump power show a further improvement compared to the times measured in Fig \ref{fig:one}d. At this point, we approach the limit of dephasing caused by hyperfine interaction. To prove this, hyperfine interaction strength can be calculated by fitting the PRC curve with the following equation:

\begin{equation}
    \bar{S}_z = S_0 \left[ \frac{1}{3} + \frac{2}{3} \frac{B_{\text{ext}}^2}{B_{\text{ext}}^2 + B_f^2} \right]
    \label{eq:nuclear_equal}
\end{equation}

\noindent where $S_0$ -- average electron-spin polarization at zero magnetic field, $B_\mathrm{ext}$ -- applied magnetic field and $B_f$ -- averaged nuclear fluctuation field strength, which is defined as the half width at half minimum (HWHM) of the PRC curve dip. However, this equation assumes that the nuclear field fluctuations are isotropic, but as we have shown, our sample shows anisotropy between z and (xy) directions. Applying the anisotropy parameter obtained from our measurements improves the fit, but the dip still does not reach the actual values. Therefore, for this, we assume that our anisotropy parameter is 0.1, which is possible due to the Hanle measurement showing significant noise and so distorting the fit values. With this assumption, the equation becomes $\bar{S}_z = S_0 \left[ 0.1 + 0.9 B_{\text{ext}}^2/(B_{\text{ext}}^2 + B_f^2) \right]$.
Fitting this equation with the PRC data shown in Fig.~\ref{fig:three} results in $B_f = 13\,$mT. From this, the spin dephasing time limited by hyperfine interaction can be calculated $T_2^* = 2\sqrt{3}\hbar / (g_e\mu_B B_f) = 3.2\,$ns, which is only slightly larger than the value we have obtained in Fig.~\ref{fig:four}a. While Fig.~\ref{fig:one}d already shows the spin dephasing saturation at low applied magnetic field, which usually indicates that the values are limited by nuclear fluctuations, this further confirms that in our sample the electron spin dephasing time is limited mainly by hyperfine interaction and not any other factors.

We were not able to provide $T_2$ measurements as we could not observe the spin mode-locking effect (SML)~\cite{Evers_2024, Greilich_2006}. One of the possible reasons is that $T_2$ is too short for SML to occur, however, seeing how an improvement in spin dynamics is observed, it would be expected that an improvement for $T_2$ would also be observed, which should make it long enough for SML to be present. Another possibility is that our signal to noise ratio was poor, which overshadowed the SML signal. Compared to MBE SK grown samples that we have previously measured, this sample exhibited increased noise and reduced signal strength. This degradation may result from point defects introduced during the stacking of multiple quantum dot layers. The sample in Ref.~\cite{Evers_2024} also has 8 layers of QDs with a density of approximately $\mathrm{10^{10}\ cm^{-2}}$, whereas our sample is 3 layers, which results in our sample having a significantly smaller effective QD density and in turn weaker signal.

\flushbottom

\section{Conclusions}

We have characterized the spin properties of InAs/InGaAs/InP quantum dots grown by droplet epitaxy using MOVPE. Our measurements reveal several key findings that demonstrate the advantages of this growth method for telecom wavelength quantum dots. Most notably, we observed a longitudinal spin relaxation time $T_1 = 2.95$\, \si{\micro\second} for electrons, representing an order of magnitude improvement over comparable MBE-grown telecom QD samples~\cite{Evers_2024, Mikhailov_2018}. While we still observe significant electron $g$-factor anisotropy, which results in greater hyperfine interaction anisotropy, we show that compared to other telecom samples, we see an improvement in anisotropy~\cite{Belykh_2016}. This enhancement is attributed to the improved structural symmetry achieved through MOVPE droplet epitaxy growth, which reduces the asymmetric growth profile common in both MOVPE and MBE strain-based growth methods~\cite{Gajjela_2022}.

\section{Acknowledgments}

The authors acknowledge the Bundesministerium für Bildung und Forschung in the frame of the project QR.N (contract no. 16KIS2201) and EPSRC Grant EP/S030751/1.

\bibliography{references} 

\end{document}